\documentclass[prb,onecolumn,12pt,showpacs,showkeys,preprintnumbers,amsmath,assume]{revtex4}
\usepackage{graphicx}
\usepackage{dcolumn}
\usepackage{bm}
\pdfoutput=1

\usepackage{natbib}
\setcitestyle{authoryear,open={},close={}}

\begin{document}

\title{An Eigenvalue Analysis of Damping in Optical Thin Plasmas}

\author{J. Fajardo$^{1,3}$, P. Contreras$^{2,3}$, and M. H. Ibañez S.$^{3}$}
\affiliation{$^{1}$ Facultad de Ciencias Básicas, Educación, Artes y Humanidades, ITSA, Barranquilla, Colombia}
\affiliation{$^{2}$ Depto de Física, Universidad de Los Andes, Mérida 5101, Venezuela}
\affiliation{$^{3}$ Centro de Física Fundamental, Universidad de Los Andes, Mérida 5101, Venezuela}
\date{\today}

\begin{abstract}

In this work, the behavior of magnetohydrodynamic waves in optically thin plasmas considering dissipative processes, thermal and magnetic diffusion, a given ionization, and the heating and cooling functions are investigated for several particular cases. A numerical eigenvalues analysis of the dimensionless secular equations according to various cases is performed for the entire set of $MHD$ equations.

\end{abstract}

\keywords{thin optical plasmas, magnetosonic waves, Alfv\`{e}n waves, thermal waves, magnetic diffusion}
\maketitle
\newpage

\section{Introduction}

In recent years analytical as well as numerical tools for working out non linear partial differential equation, in particular, those governing general fluids have been enormously improved, nonetheless the linear problem resulting from analyzing these equations remains to be very important for many reasons: 

\begin{itemize}
    \item The associated eigenvalue problem describes the behavior of magnetohydrodynamic waves (MHD) and other waves, say for instance thermal and radiation waves. 
    \item Understanding the behavior of linear waves allows to understand many physical aspects of non linear problems, for instance the onset of the turbulence as well as its closed relation with it (\cite{De05,Bu06,He06,Pa09,Br12}).
    \item The linear approach is closely related to the problem of stability of different flows and gas structures in different physical fields, in particular in astrophysical problems: planetary atmospheres, Earth oceans, stellar interiors (\cite{He06,Pa79,He05}) stellar atmospheres, say for instance the solar atmosphere, interstellar medium, intracluster media (\cite{Ke76,Ki91}).
\end{itemize}

However, the present work is limited to the analysis of some aspects of MHD wave propagation in optically thin plasmas of interest in astrophysics. This limited problem however is of great importance and has been extensively worked out in:

\begin{itemize}

    \item The solar atmosphere (\cite{Ib85,Ib92,Ib93,Da88,Fi65,Gi79,La60,Mi78,Ro78,Sp62,St92,Ma86,Ve79,Ku06,Ro00}).
    
    \item The interstellar and intracluster media 
    (\cite{Fi65,Ib04,Go70,Mc77}).
    
\end{itemize}
 
There are several aspects which have not been considered previously in the Alfv\`{e}n wave damping analysis and in the magnetosonic wave analysis and the associated eigenvalue problem for optically thin plasmas, as will be seen and discussed at the present work.

In section II, the set of MHD equations is linearized, leading to two independent cases where each matrix generates a dispersion relation whose roots for the case of Alfv\`{e}n waves are a complex equation. This relationship led to a deduction of the Landau damping expression in a different way to the one presented by classical texts.

In section III, for the linear approximation both modes are studied for the thermal and magneto-acoustic cases. They are damped by thermal conduction, viscosity and the influence of the cooling-heating function. The complex eigen-equation is described for several asymptotic cases:
\begin{itemize}
    \item  The case when only one dissipative process is taken into account.
    \item The case with only magnetic diffusion term ($\tilde{\nu}_{m}$).
\end{itemize}

In section IV, in the energy equation the dissipative terms were neglected, but the effects of the heat/loss were accounted, because of its great importance in many astrophysical as well as laboratory plasma applications. 

Finally, in section V, the kinetic coefficients in a magnetic field, for the case of a recombining hydrogen plasma are discussed.

\section{General set of magnetohydrodynamic equations}

If dissipative effects are accounted for a recombining gas, for an an optically thin and heat conducting plasma, the well known basic MHD equations can be written as (\cite{La60,St92,Ib04,Fa16})

\begin{equation}
\frac{\partial H_{k}}{\partial x_{k}}=0~,  \label{G}
\end{equation}

\begin{equation}
\frac{\partial H_{i}}{\partial t}=-\epsilon_{ijk}\frac{\partial }{\partial
x_{j}}\left( v_{j}H_{k}-v_{k}H_{j}\right) \mathbf{+}\frac{c^{2}}{4\pi \sigma 
}\frac{\partial ^{2}H_{i}}{\partial x_{k}^{2}}~,  \label{H}
\end{equation}

\begin{equation}
\frac{\mathbf{\partial }\rho }{\mathbf{\partial }t}+\frac{\partial (\rho
v_{i})}{\partial x_{i}}=0~,  \label{Ma}
\end{equation}

\begin{equation}
\frac{d\xi }{dt}+X(\rho ,T,\xi )=0,  \label{Xrate}
\end{equation}

\begin{equation}
\rho \frac{dv_{i}}{dt}\mathbf{=-}\frac{\partial p}{\partial x_{i}}\mathbf{+}
\frac{1}{4\pi }\frac{\partial }{\partial x_{k}}\left( H_{i}H_{k}-\frac{1}{2}
H^{2}\delta _{ik}\right) +\frac{\partial \sigma 
{\acute{}}
_{ik}}{\partial x_{k}}~,  \label{motion}
\end{equation}

\begin{equation}
\rho T\frac{ds}{dt}=-\rho L(\rho ,T,\xi )+\frac{\partial }{\partial x_{i}}(\kappa _{ik}\frac{\partial T}{\partial x_{k}})+\frac{c^{2}}{16\pi
^{2}\sigma }\left[ -\epsilon _{jki}\frac{\partial H_{j}}{\partial x_{k}}\right] ^{2}+\sigma 
{\acute{}}_{ik}\frac{\partial v_{i}}{\partial x_{k}}~,  \label{energy}
\end{equation}

and, 
 
\begin{equation}
p=\frac{N_{0}k_{B}}{\mu(\xi )}\rho T, \label{state}
\end{equation}

where $H_{i}$ and $v_{i}$ are the $i$\emph{-esime} components of the
magnetic field and velocity, respectively. $\epsilon_{jki}$ is the
permutation symbol $\delta_{ik}$ is Kronecker delta symbol, $c$ the light speed and $\rho$, $p$, $T$, $\xi $, $c_{v}$,$N_{0}$, $k_{B}$ and $\mu(\xi)$ respectively are mass density, pressure, temperature, ionization degree, specific heat at constant volume, the Avogadro number, the Boltzmann constant and the mean molecular weight of the gas. 

$X(\rho ,T,\xi )$ is the net ionization rate and $L(\rho ,T,\xi )$ is the heat-loss function defined as energy losses minus energy gains per unit mass and time, which can be written as
$L(\rho ,T,\xi ) =$ $L (\rho, T, \xi)_{output} - L (\rho, T,\xi)_{input}$.

Additionally, $\kappa_{ik}$ and $\sigma_{ik}$ are the thermal conduction and the viscous stress tensor, respectively. 

The thermal conduction coefficient $\kappa_{ik}$ generally is weakly dependent on density but strongly dependent on temperature (\cite{Pa53,Sp62,Br65,Ib16}).

Strictly speaking the induction equation becomes rather complicated, in particular, the electrical conductivity $\sigma $ is also a
tensor, however, for sake of simplicity and taking into account that $\sigma_{\parallel}/\sigma_{\perp}=1.96$, this quantity will be assumed as a scalar of magnitude $\sigma$ and the induction equation will be assumed in the simplified form given by Eq.(\ref{H}) (\cite{Br65}). 

This set of equations reduces to the known MHD equations (\cite{La60}) when the heat/loss term is neglected.

\section{Eigenvalue analysis of the type of magneto hydrodynamic waves}

For an inert plasma if all dissipative processes are neglected the Eqs. (\ref{G}), (\ref{Ma}) and (\ref{state}) hold and the Eqs. (\ref{H})-(\ref{energy}) simplify,i.e the set of ideal MHD equations can be written as (\cite{Fa16}

\begin{equation}
\frac{\partial \mathbf{H}}{\partial t}=\mathbf{\nabla \times (v\times H)},
\label{H0}
\end{equation}

\begin{equation}
\frac{\partial \rho }{\partial t}+\mathbf{\nabla \cdot (}\rho \mathbf{v)=0},
\label{M0}
\end{equation}

\begin{equation}
\frac{d\mathbf{v}}{dt}\mathbf{=-}\frac{1}{\rho }\mathbf{\nabla }p\mathbf{-}
\frac{1}{4\pi \rho }\mathbf{H\times curl(H)}~,  \label{Mm0}
\end{equation}

\begin{equation}
\frac{\mathbf{\partial }s}{\mathbf{\partial }t}+\mathbf{v\cdot \nabla }s=0.
\label{entropy0}
\end{equation}

For small disturbances superposed to an steady flow with velocity $\mathbf{V}_{0}$, magnetic field $\mathbf{H}_{0}$, pressure $p_{0}$ and mass density $\rho_{0}$ (\cite{La60,Ib09})

\[
\mathbf{v=V}_{0}+\mathbf{v}^{\prime },\mathbf{\quad H=H}_{0}+\mathbf{h,\quad 
}p=p_{0}+p^{\prime }\mathbf{,\quad }\rho =\rho _{0}+\rho ^{\prime }
\]

Where $\mathbf{v}^{\prime },\mathbf{h,}$ $p^{\prime}$ and 
$\rho^{\prime}$ are functions of ($x,y,z,t$). Therefore, Eqs. (\ref{H0})-(\ref{entropy0}), up to the first order, become

\begin{equation}
\mathbf{\nabla \cdot h}=0, \label{G1}
\end{equation}

\begin{equation}
\frac{\partial \mathbf{h}}{\partial t}-\mathbf{\nabla \times (v}^{\prime }
\mathbf{\times H}_{0}\mathbf{)-\nabla \times (V}_{0}\mathbf{\times h)}=0,
\label{h1}
\end{equation}

\begin{equation}
\frac{\mathbf{\partial }s^{\prime}}{\mathbf{\partial }t}+\mathbf{V}_{0}
\mathbf{\cdot \nabla }s^{\prime}=0,  \label{s1}
\end{equation}

\begin{equation}
\frac{\partial p^{\prime }}{\partial t}+\mathbf{V}_{0}\mathbf{\cdot \nabla} p^{\prime }+\rho _{0}u_{0}^{2}\mathbf{\nabla \cdot \mathbf{v}^{\prime }}=0, \label{ro1}
\end{equation}

\begin{equation}
\frac{\partial \mathbf{v}^{\prime }}{\partial t}+(\mathbf{V}_{0}\mathbf{
\cdot \nabla )\mathbf{v}^{\prime }+}\frac{1}{\rho _{0}}\mathbf{\nabla }
p^{\prime }+\frac{1}{4\pi \rho _{0}}\mathbf{H}_{0}\mathbf{\times curl(h)}, \label{Mm1}
\end{equation}

Furthermore, for disturbances of the form 
$\sim \exp [i(\mathbf{k\cdot r-} \omega t)]$, Eqs. (\ref{G1})-(\ref{Mm1}) reduce to

\begin{equation}
\mathbf{k\cdot h}=0, \label{lim1}
\end{equation}

\begin{equation}
-\omega \mathbf{h-k\times (v}^{\prime }\times \mathbf{H}_{0}\mathbf{
)-k\times (V_{0}\times h)}=0~, \label{lim2}
\end{equation}

\begin{equation}
\left( \mathbf{V}_{0}\mathbf{\cdot k}-\omega \right) s^{\prime }=0~,
\label{li1}
\end{equation}

\begin{equation}
\left( \mathbf{V}_{0}\mathbf{\cdot k}-\omega \right) p^{\prime }+\rho
_{0}u_{0}^{2}\mathbf{k\cdot v}^{\prime }=0~,  \label{lid2}
\end{equation}

\begin{equation}
\left( \mathbf{V_{0}\cdot k}-\omega \right) \mathbf{v}^{\prime }+\frac{1}{
\rho _{0}}p^{\prime }\mathbf{k}+\frac{1}{4\pi \rho _{0}}\mathbf{H}_{0}
\mathbf{\times (k\times h)}=0~,  \label{lid3}
\end{equation}

where $\rho ^{\prime }=p^{\prime }/u_{0}^{2}+(\partial \rho /\partial s)_{p}s^{\prime }$ and $u_{0}^{2}=(\partial p/\partial \rho )_{s}$. 

Eq. (\ref{lim1}) implies that $\mathbf{h}$ is perpendicular to $\mathbf{k}$, therefore, from Eq. (\ref{lid3}) follows that $p^{\prime }=0$ and Eqs. (\ref{li1}) and (\ref{lid2}) \ reduce to

\begin{equation}
\left( \mathbf{V}_{0}\mathbf{\cdot k}-\omega \right) =0,\quad s^{\prime}\neq 0,\quad \mathbf{k\times v}^{\prime }\neq 0~.  \label{evw1}
\end{equation}

Without lost of generality, $\mathbf{V}_{0}$ and $\mathbf{H}_{0}$\ are assumed to be on the $x\mathcal{-}y$ plane. The above relations define an entropy vortex wave which is carried along with the flow and is independent on other linear modes which correspond to the solutions

\[
\left( \mathbf{V}_{0}\mathbf{\cdot k}-\omega \right) \neq 0,\quad s^{\prime
}=0,\quad \mathbf{k\cdot v}^{\prime }=0.
\]

These modes are defined by the eigen equations
\begin{equation}
\left( 
\begin{array}{cc}
u-V_{x} & H_{x} \\ 
\frac{H_{x}}{4\pi \rho } & u
\end{array}
\right) \left( 
\begin{array}{c}
h_{z} \\ 
v_{z}
\end{array}
\right) =0,  \label{Alf0}
\end{equation}
and 

\begin{equation}
\left( 
\begin{array}{ccc}
u-V_{y} & -H_{y} & H_{x} \\ 
\frac{H_{x}}{4\pi \rho } & 0 & u \\ 
-\frac{H_{y}}{4\pi \rho } & (u-V_{x})-\frac{u_{0}^{2}}{u-V_{x}} & 0%
\end{array}
\right) \left( 
\begin{array}{c}
h_{y} \\ 
v_{x} \\ 
v_{y}
\end{array}
\right) =0,  \label{mhd0}
\end{equation}

where $u=\omega /k$ is the $\emph{phase}$ $\emph{velocity.}$ Herein after, the sub index $_{0}$ indicating equilibrium values will be omitted, except for $u_{0}$, that is, the adiabatic sound speed. 
$\mathbf{k}$ is taken here to be along the $x$-axis.

In the particular case of a plasma initially at rest, the compatibility condition for the Equations (\ref{Alf0}) and (\ref{mhd0}) respectively become

\begin{equation}
u^{2}=\frac{H_{x}^{2}}{4\pi \rho},  \label{Alf1}
\end{equation}
and

\begin{equation}
u^{4}-\left( \frac{H^{2}}{4\pi \,\rho }+u_{0}^{2}\right) u^{2}+\frac{H_{x}^{2}}{4\pi \rho }u_{0}^{2}=0~,  \label{mhd1}
\end{equation}

As it is well known, Eqs. (\ref{Alf0} and (\ref{Alf1}) define the Alfv\`{e}n modes, and Eqs. (\ref{mhd0}) and (\ref{mhd1}) define the fast and slow magnetosonic modes (Alfv\`{e}n suggested the existence of hydro-magnetic waves in 1942, \cite{La60,Ib09}).

In the general case of a plasma flowing with an initial constant velocity $\mathbf{V}_{0}$ the dispersion relations are modified accordingly but the nature of the wave modes remains.

In conclusion, as far as the linear approximation concerns, there are three kind of waves in a plasma flow, and which are independent each other: 

\begin{itemize}
    \item The entropy-vortex modes.
    \item The Alfv\`{e}n modes.
    \item The magnetosonic waves.
\end{itemize}

Te entropy entropy-vortex modes were worked out in (\cite{Ib16,ibl})

\section{Dissipative processes in magneto hydrodynamic waves with a given ionization and heat/loss effects}

For a plasma with a given ionization and taking into account dissipative and heat/loss effects, the linearization of Eqs. (\ref{G})-(\ref{state}) give, as in the ideal case, two sets of equations independent each other, that is (\cite{Ib11})

\begin{equation}
\left( 
\begin{array}{cc}
\omega +i\frac{c^{2}k^{2}}{4\pi \sigma } & H_{x}k \\ 
\frac{H_{x}k}{4\pi \rho } & \omega +i\frac{\eta k^{2}}{\rho }
\end{array}
\right) \left( 
\begin{array}{c}
h_{z} \\ 
v_{z}
\end{array}
\right) =0  \label{Alfven}
\end{equation}

\begin{equation}
\left( 
\begin{array}{cccc}
\omega (-i\omega +\Omega ) & 0 & \left( iu_{0}^{2}\rho \omega +\Gamma
\right) k & 0 \\ 
0 & \omega +i\frac{c^{2}k^{2}}{4\pi \sigma } & -kH_{y} & kH_{x} \\ 
-k & -\frac{H_{y}k}{4\pi } & \rho \omega +i\left( \frac{4\eta }{3}+\zeta
\right) k^{2} & 0 \\ 
0 & \frac{H_{x}k}{4\pi } & 0 & \rho \omega +i\eta k^{2}
\end{array}
\right) \left( 
\begin{array}{c}
p^{\prime } \\ 
h_{y} \\ 
v_{x} \\ 
v_{y}
\end{array}
\right) =0  \label{madw}
\end{equation}
where

\begin{equation}
\Omega =\frac{1}{c_{v}}\left( \frac{\kappa k^{2}}{\rho }+L_{T}\right)
,~\Gamma =\rho (\gamma -1)\left[ \rho L_{\rho }-T\left( \frac{\kappa k^{2}}{\rho }+L_{T}\right) \right] ~.  \label{11a13}
\end{equation}

The coefficients of viscosity appearing into the viscous stress tensor $\sigma _{ij}$ are tensors due to the anisotropy introduced by the magnetic field, the ratio between the parallel and perpendicular kinematic viscosity becomes $\eta _{\parallel}/\eta _{\perp }\approx 1.98$ (\cite{Sp62}), therefore, this coefficient as well as the bulk viscosity $\zeta$ can be assumed as scalars of magnitude $\eta$ and $\zeta$, respectively, in the equations (\ref{Alfven}) and (\ref{madw}). 

Additionally, the strong anisotropy inherent in the thermal conduction tensor $\kappa _{ij}$\ ($\eta_{\perp }/\eta _{\parallel }\approx 10^{-12}$) has been taken into account assuming the heat flux vector to be 

\begin{equation}
\mathbf{q=}-\left( \kappa _{\parallel }\frac{\partial T}{\partial
s_{\parallel }}\mathbf{n}_{\parallel }+\kappa _{\perp }\frac{\partial T}{\partial s_{\perp }}\mathbf{n}_{\perp }\right) ~,  \label{heatF}
\end{equation}

where $\mathbf{n}_{\parallel}$ and $\mathbf{n}_{\perp}$ are unit vectors along and perpendicular to $\mathbf{H}_{0}$, respectively. 

Therefore,

\begin{equation}
\kappa =\left[ \kappa _{\parallel}\cos ^{2}\theta +\kappa_{\perp }\sin^{2}\theta \right],  \label{kappa}
\end{equation}

where $\theta =\cos^{-1}\left(H_{x}/H_{0}\right)$.

In dimensionless form Eq. (\ref{madw}) can be written as

\begin{equation}
\left( 
\begin{array}{cccc}
1+i(\tilde{\kappa}\tilde{k}^{2}+\tilde{L}_{T}) & 0 & -1+i\gamma ^{-1}(\tilde{L}_{\rho }-\tilde{L}_{T}-\tilde{\kappa}\tilde{k}^{2})\tilde{k} & 0 \\ 
0 & 1+i\tilde{\nu}_{m}\tilde{k}^{2} & -\sin (\theta )\tilde{k} & \cos(\theta )\tilde{k} \\ 
-\beta ^{2}\tilde{k} & -\sin (\theta )\tilde{k} & 1+i\left( \frac{4}{3}
\tilde{\nu}+\tilde{\nu}_{b}\right) \tilde{k}^{2} & 0 \\ 
0 & \cos (\theta )\tilde{k} & 0 & 1+i\tilde{\nu}\tilde{k}^{2}%
\end{array}
\right) \left( 
\begin{array}{c}
\tilde{p}^{\prime } \\ 
\tilde{h}_{y} \\ 
\tilde{v}_{x} \\ 
\tilde{v}_{y}
\end{array}
\right) =0  \label{amadw}
\end{equation}
where 

\begin{equation}
\tilde{\kappa} = \frac{\kappa_{\parallel} \omega}{\rho c_{v}a^{2}} 
\left[\cos^{2}(\theta)+ \frac{\kappa_{\perp} }{\kappa_{\parallel}} \sin^{2}(\theta)\right],
\label{akappa}
\end{equation}

and $\tilde{k}=$ $a k/\omega$, $a=H_{0}/\sqrt{4\pi \rho }$, $\tilde{L}_{T}=$ $L_{T}/c_{v}\omega$, $\tilde{L}_{\rho }=$ $\rho L_{\rho }/c_{v}T\omega $, $\tilde{\nu}=\omega \eta /\rho a^{2}$, $\tilde{\nu}_{b}=\omega \zeta /\rho a^{2}$, $\tilde{\nu}_{m}=\omega c^{2}/4\pi \sigma a^{2}$, $\beta =u_{0}/a$, $\tilde{p}=$ $p/\rho u_{0}^{2}$, $\tilde{h}_{y}=$ $h_{y}/H_{0}$, $\tilde{v}_{x}=$ $v_{x}/a$, and $\tilde{v}_{y}=v_{y}/a$.

\subsection{Numerical results for the Alfv\`{e}n wave damping}

The corresponding dimensionless secular equation of the system of equations (\ref{Alfven}) becomes equal to

\begin{equation}
\tilde{\nu}\tilde{\nu}_{m}\tilde{k}^{4}+\left[ 1-i\left( \tilde{\nu}+\tilde{\nu}_{m}\right) \right] \tilde{k}^{2}-1=0~,  \label{Alfad}
\end{equation}

where $\tilde{\nu}=\omega \eta /\rho a_{x}^{2}$, $\tilde{\nu}_{m}=\omega c^{2}/4\pi \sigma a_{x}^{2}$, $\,\,\tilde{k}=a_{x}k/\omega $ and $a_{x}=\left\vert H_{x}\right\vert /\sqrt{4\pi \rho }$.

The roots of Eq. (\ref{Alfad}) are complex, that is, $\tilde{k}=\tilde{k}_{r}+i\tilde{k}_{i}$ where $\tilde{k}_{r}$ and $\tilde{k}_{i}$ are real quantities.

Due to the fact that $\tilde{\nu}$ as well as $\tilde{\nu}_{m}$ ($\approx \bar{l}/\lambda \ll 1$, where $\bar{l}$ is the mean free path and $\lambda$ the Alfv\`{e}n wave length (\cite{La60}), the quartic term of Eq. (\ref{Alfad}) can be neglected and the resulting quadratic equation has the solution sought.

\begin{equation}
\tilde{k}_{i}\approx \frac{1}{2}\left( \tilde{\nu}+\tilde{\nu}_{m}\right).
\label{AlfLL}
\end{equation}

Because the disturbance has been taken in the form $\sim \exp (\mathbf{k\cdot r-}i\omega t)$, therefore $k_{i}=\omega \tilde{k}_{i}/a_{x}$ becomes the absorption coefficient. 

One must remark that the expression (\ref{AlfLL}) holds as far as the damping per wave length is very small.  This expression was first obtained by (\cite{La60}) in a different way. 

Strictly speaking if both coefficients $\tilde{\nu}$ and $\tilde{\nu}_{m}$ are different from $\emph{zero}$, Eq.(\ref{Alfad}) has two roots for $\tilde{k}^{2}$, however only one of the roots fulfils the condition $\tilde{k}_{i} \ll 1$ for which the present approximation holds. 

For this physical meaningful mode, the velocities ($v$), the damping coefficient ($k_{i}$), the damping per unit wave length ($l_{d}/\lambda =k_{r}/2\pi k_{i}$) and the ratio $\left\vert h_{z}/v_{z}\sqrt{\pi \rho }\right\vert $ have been plotted in Fig 1 as functions of $\tilde{\nu}$ for three different values of of the ratio $\nu _{m}/\nu$ ($=0.1$ blue line, $1$ black line and $10$ red line).

Additionally, for the ($k_{i}$), the damping per unit wave length, the Landau approximation for (\ref{AlfLL}) has been plotted (3 pointed lines in Fig. 1b).

\begin{figure}[ht]
\includegraphics[width = 5.0 in, height= 4.5 in]{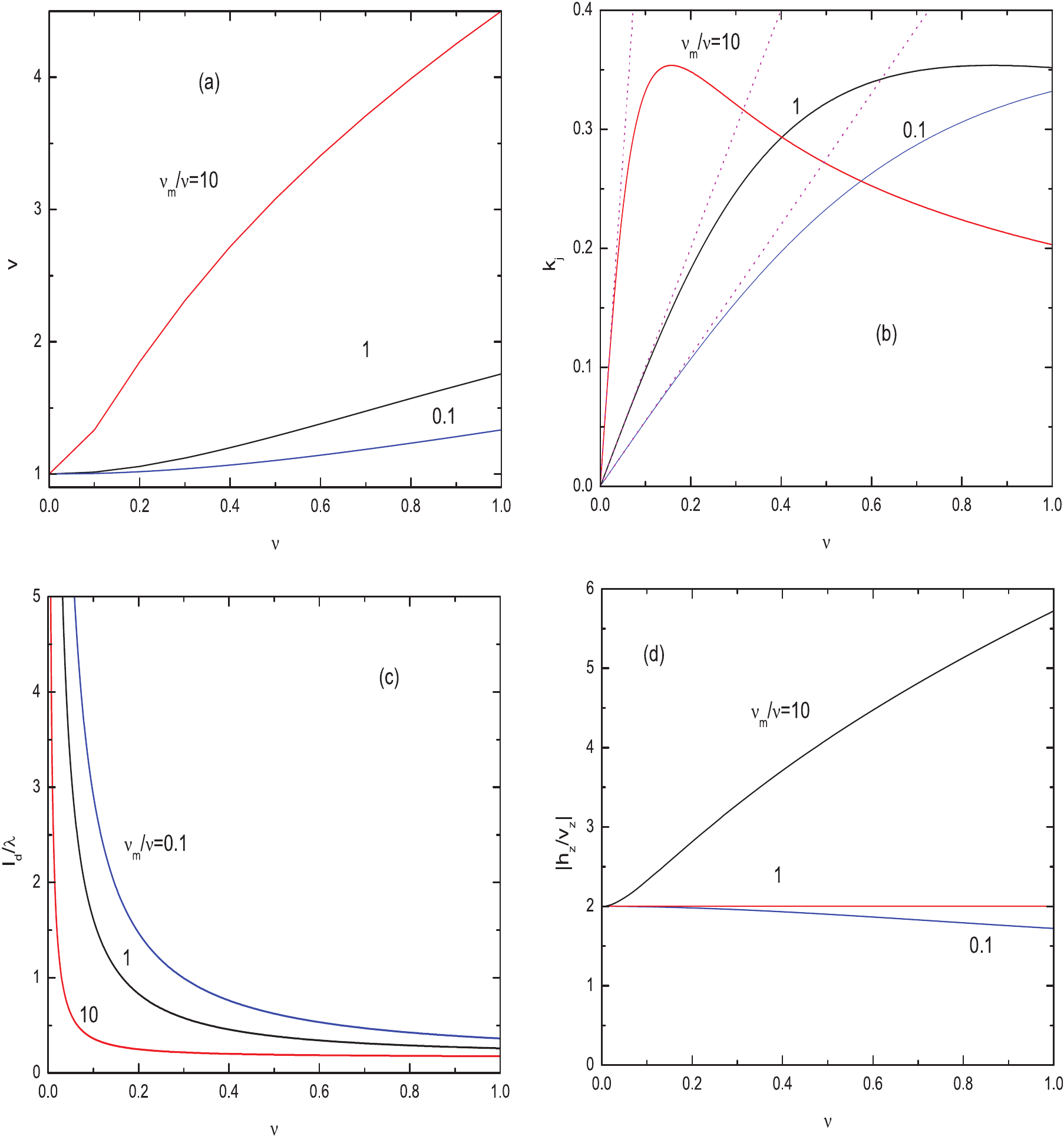}
\caption{The velocity modes ($v$) (a), the damping coefficient ($k_{i}$) (b), the damping per unit wave length ($l_{d}/\lambda =k_{r}/2\pi k_{i}$) (c), and the ratio $\left\vert h_{z}/v_{z}\sqrt{\pi\rho }\right\vert$ (d) have been plotted in Fig 1 as functions of $\tilde{\nu}$ for three different values of of the ratio $\nu _{m}/\nu$ ($=0.1$ blue line, $1$ black line and $10$ red line)}
\end{figure}

\clearpage

\subsection{Numerical results for the magnetosonic and thermal waves}

The condition of compatibility of the system of equations (\ref{amadw}) can be written as

\begin{equation}
\left( a_{0r} + i \; a_{0i}\right) \tilde{k}^{8}+\left( a_{1r}+i \; a_{1i}\right) \tilde{k}^{6}+\left( a_{2r}+ i \; a_{2i}\right) \tilde{k}^{4}+\left(a_{3r}+i \; a_{3i}\right) \tilde{k}^{2}+1+i\tilde{L}_{T}=0,  \label{madis}
\end{equation}
the coefficients $a_{kl}$ are defined as

\[
a_{0r}=\tilde{\kappa} \; \tilde{\nu} \; \tilde{\nu}_{m} \; \left( \frac{4}{3}\tilde{\nu} + \tilde{\nu}_{b}\right),
\]

\[
a_{0i}= {\beta }^{2} \; \gamma ^{-1} \; \tilde{\kappa} \; \tilde{\nu} \; \tilde{\nu}_{m},
\]

\[
a_{1r} = {\beta }^{2} \; \gamma ^{-1} \; \tilde{\kappa}\left( \tilde{\nu}+\tilde{\nu}
_{m}\right) +\tilde{\kappa}\left[ \tilde{\nu}\left( 1+\frac{1}{3}\cos^{2}\left( \theta \right) \right) +\tilde{\nu}_{b}\cos ^{2}\left( \theta
\right) \right] +\tilde{\nu}\tilde{\nu}_{m}\left[{\beta }^{2}+\left( \frac{4}{3\,}{\tilde{\nu}+}\tilde{\nu}_{b}\right) \tilde{L}_{T}\right],
\]

\[
a_{1i}= {\beta }^{2} \; \gamma ^{-1} \; \left[\tilde{\kappa}\cos ^{2}\left( \theta \right) +\tilde{\nu}\tilde{\nu}_{m}\left( \tilde{L}_{T} - \tilde{L}_{\rho }\right) \right] -\tilde{\kappa}\left[ \tilde{\nu}\left( \frac{4}{3}\; {\tilde{\nu}+}\tilde{\nu}_{b}+\frac{7}{3}\tilde{\nu}_{m}\right) +\tilde{\nu}_{b} \tilde{\nu}_{m}\right] - {\tilde{\nu}}\left( \frac{4}{3}{\tilde{\nu}}+\tilde{\nu}_{b}\right) \tilde{\nu}_{m},
\]

\[
a_{2r}={\beta }^{2}\gamma ^{-1}\left( \tilde{\nu} \,+\tilde{\nu} _{m}\,\right) \left(
\tilde{L}{T}-\tilde{L}{\rho }\right) -\tilde{\kappa} \left( \frac{7}{3}\,\tilde{\nu} +\tilde{\nu} _{b}+\tilde{\nu}_{m}\right) +
\]

\[
 \left[ \tilde{\nu} \left( 1+\,\frac{1}{3}\cos ^{2}\left( \theta \right)
\right) +\tilde{\nu} {b}\cos ^{2}\left( \theta \right) \right] \tilde{L}{T}
   -\left[ \tilde{\nu}
\left(\frac{4}{3}{\tilde{\nu}}+\tilde{\nu} _{b}\,+\,\frac{7}{3}\tilde{\nu} _{m}\,\right) +\tilde{\nu}
_{b}\tilde{\nu}_{m}\,\right] +{\beta }^{2}\cos^{2}\left( \theta \right),
\]

\[
a_{2i}={\beta }^{2}\gamma ^{-1}\left[ \left( \tilde{L}{T}-L{\rho }\right) \cos
^{2}\left( \theta \right) -\tilde{\kappa} \right] -\tilde{\kappa} -\left[ \tilde{\nu} \left( \frac{4
}{3}\,{\tilde{\nu} +}\tilde{\nu} _{b}\,+\frac{7}{3}\,\tilde{\nu} _{m}\,\right) +\tilde{\nu} _{m}\,\tilde{\nu} _{b}
\right] \tilde{L}_{T} -
\]

\[
\left[ \tilde{\nu} \left( {\beta }^{2}+1+\frac{1}{3}\cos ^{2}\left(
\theta \right) \right) +\tilde{\nu} _{b}\cos ^{2}\left( \theta \right) +{\beta }
^{2}\tilde{\nu} _{m}\right] ~,
\]

\[
a_{3r}=-\left[ \left( \frac{7}{3}\tilde{\nu}+\tilde{\nu}_{b}+\tilde{\nu}
_{m}\right) \tilde{L}_{T}+1+{\beta }^{2}\right],
\]
and

\begin{equation}
a_{3i}={\beta }^{2}\gamma ^{-1}\left( \tilde{L}_{\rho }-\tilde{L}_{T}\right)
+\tilde{\kappa}-\tilde{L}_{T}+\frac{7}{3}\,\tilde{\nu}+\tilde{\nu}_{b} + \tilde{\nu}_{m}.  \label{coedisper}
\end{equation}

Generally speaking, the parameters defining the coefficients of the fourth order polynomial in $\tilde{k}^{2}$ (\ref{madis}) depend on two thermodynamic quantities, say $\rho$ and $T$ and two quantities defining the magnetic field, i.e. $H$ and $\theta$. Therefore, these parameters define the corresponding four wave modes resulting from the equation (\ref{madis}). 

The square root $\pm \tilde{k}$ represents two waves propagating in opposite directions each other. The angle ranges between $0\leq \theta \leq $ $\pi/2$, but the ranges for $\rho $, $T$ and $H$ where the dispersion relation (\ref{madis}) holds is rather wide. 

Therefore, here only a few asymptotic cases will be
discussed and the solution of the full polynomial (\ref{madis}) will serve only for specific applications.

If all dissipative mechanisms as well as the heat/input effects are
neglected and $\theta \neq \frac{\pi}{2}$, the dispersion relation reduces to a quadratic polynomial for $\tilde{k}^{2}$ (\ref{mhd1}) corresponding to the undamped fast and slow magnetosonic waves $(mw)$ (\cite{La60}) but when $\theta = \frac{\pi}{2}$ only the fast magnetosonic mode remains.

If the only one dissipative process taken into account is the thermal conductivity and $\theta \neq \frac{\pi}{2}$, Eq.(\ref{madis}) reduces to a cubic polynomial the roots of which correspond to two damped magnetosonic waves $sw$ and a thermal wave $Thw$. 

When $\theta =0$ a root becomes $\tilde{k}=1$ for which $\tilde{p}^{\prime }=0$ and $\tilde{v}_{x}=0$, corresponding to an undamped Alfv\`{e}n wave $Aw$ with values of $\left\vert \tilde{h}_{y}/\tilde{v}_{y}\right\vert =1$. 

The another two roots with $\tilde{k}\neq 1$ are a damped magnetosonic wave $sw$ and an over damped thermal wave $Thw$ for which $p =\left\vert \rho \omega v_{x}/k\right\vert$, all of which all plotted in Fig. 2.  

In Fig. 2, the phase velocity (a), the damping coefficient (b) and the damping per unit wave length (c) are plotted for three different values of $\beta =0.2$ (red lines), $1$ (blue lines), $2$ (green lines) as function of $\kappa_{0}$ . 

\begin{figure}[ht]
\includegraphics[width = 6.0 in, height= 5.5 in]{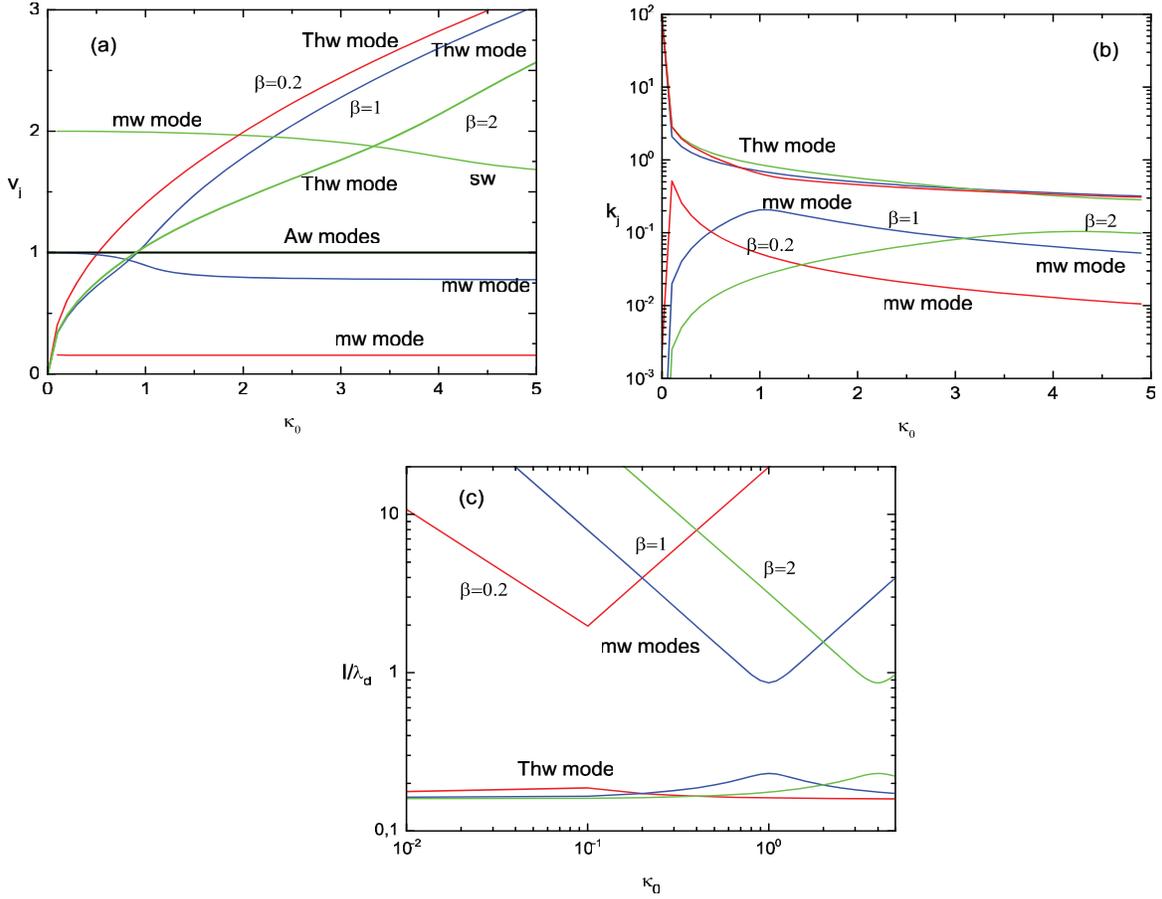}
\caption{For $\theta \neq \frac{\pi}{2}$ case the phase velocity (a) is plotted for three different values of $\beta =0.2$ (red lines), $1$ (blue lines), $2$ (green lines) are shown in the case of a thermal mode $Thw$ and the $sw$ mode, also the undamped Alfv\`{e}n mode $Aw$ is the solid black line for the three cases. The damping coefficient (b) and the damping per unit wave length (c) are plotted for three different values of $\beta =0.2$ (red lines), $1$ (blue lines), $2$ (green lines).}
\end{figure}

Note that the maximum damping of the magnetosonic wave (red $mw$ line, occurs at the same value of $\kappa_{0}$ at which the maximum damping of the thermal wave occurs for the three $\beta$ $Thw$ values in Fig. 2(b) (\cite{Ib85,Ib93}).

If $\theta =\frac{\pi}{2}$ the dispersion equation reduces to a quadratic equation, one root becomes a damped thermal wave and the another one a damped magnetosonic wave for which $\tilde{v}_{y}=0$, \ $\left\vert \tilde{h}_{y}/\tilde{v}_{x}\right\vert =\left\vert \tilde{k}\right\vert$, and $\left\vert \tilde{p} /\tilde{v}_{x}\right\vert =\beta ^{2}\left\vert (1-\tilde{k}^{2})/\tilde{k} \right\vert$, see Fig. 3 where the above two wave modes are plotted for $\beta =0.2$ (red lines), $\beta=1$ (blue lines), $\beta =2$ (green lines) for both $mw$ and $Thw$ modes. 

\begin{figure}[ht]
\includegraphics[width = 6.0 in, height= 5.5 in]{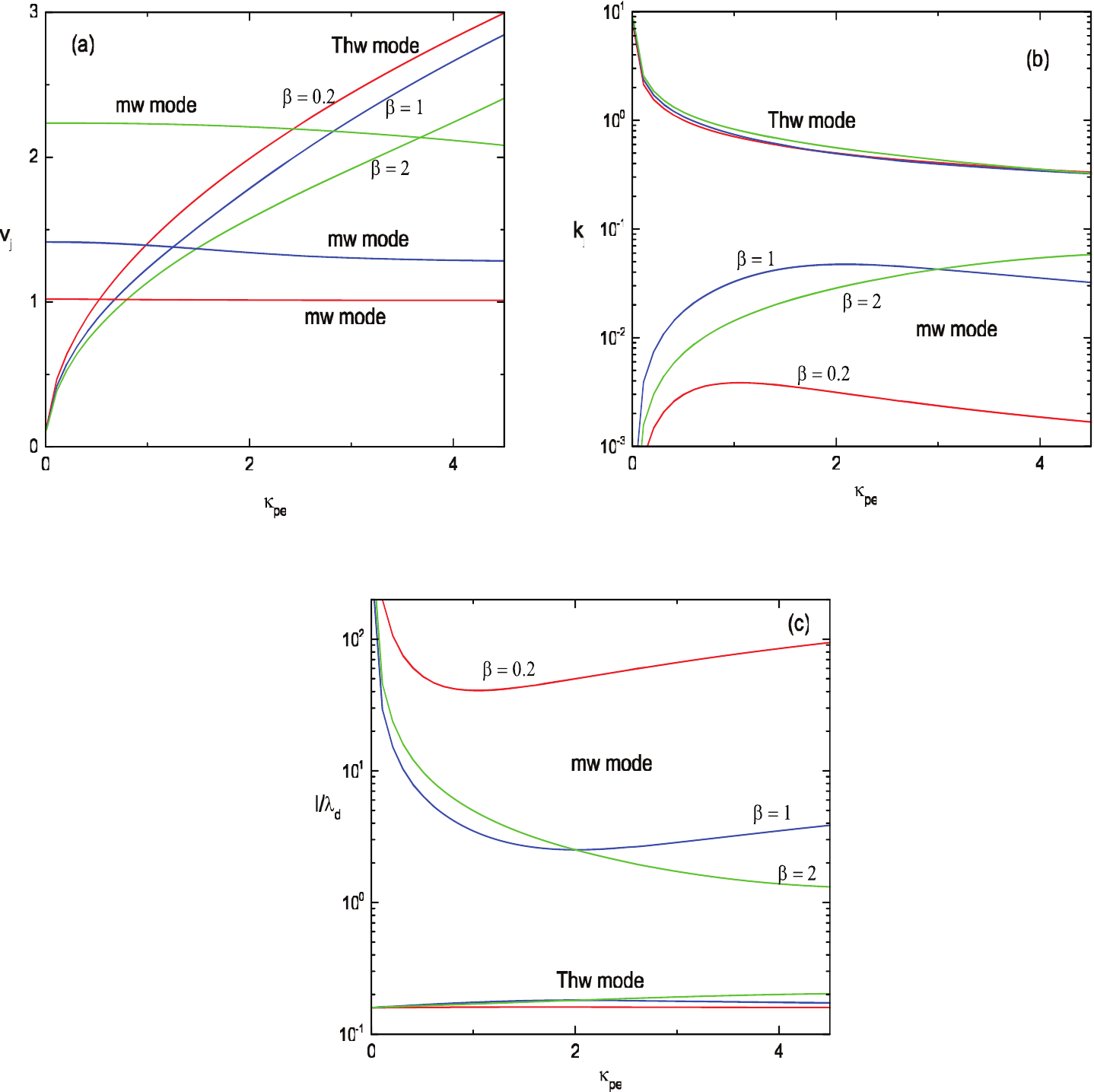}
\caption{For the dispersion equation (\ref{madis}) using the thermal conductivity with $\mathbf{\theta =\pi/2}$, the two wave modes $mw$ and $Thw$ are plotted for $\beta =0.2$ (red lines), $\beta=1$ (blue lines), $\beta =2$ (green lines). In this figure however, the wave parameters have been plotted as function of $\tilde{\kappa}_{\perp}=\left(\kappa_{\perp}\omega/\rho c_{v}a^{2}\right)$ instead of $\tilde{\kappa}_{\parallel}$.}
\end{figure}

Here one must emphasize that in the above figures the wave parameters have been plotted as function of 
$\tilde{\kappa}_{\perp }=\left( \kappa _{\perp }\omega /\rho c_{v}a^{2}\right)$ instead of $\tilde{\kappa}_{\parallel}$, 
i.e. the scales involved here are quite different (by a factor of the order of $10^{12}$) from those involved in Fig. 2.

For an angle $\theta \neq 0$ and $\theta $ $\neq \pi /2$ there are
three modes, one thermal and two magnetosonic waves (the fast and slow) modes, for which the amplitudes are related by

\begin{equation}
\left\vert \frac{\tilde{h}_{y}}{\tilde{v}_{y}}\right\vert =\frac{1}{
\left\vert \tilde{k}\cos (\theta )\right\vert }~,\qquad \left\vert \frac{
\tilde{v}_{x}}{\tilde{v}_{y}}\right\vert =\frac{1}{\left\vert \tilde{k}^{2}
\left[ \sin (\theta )-\cos (\theta )\right] \cos (\theta )\right\vert},\qquad \left\vert \frac{\tilde{p}
{\acute{}}
}{\tilde{v}_{y}}\right\vert =\left\vert \frac{1+\gamma ^{-1}\tilde{\kappa}
\tilde{k}^{3}}{1+\tilde{\kappa}\tilde{k}^{2}}\right\vert \left\vert \frac{\tilde{v}_{x}}{\tilde{v}_{y}}\right\vert,  \label{amplite0}
\end{equation}

The Fig.4 corresponds to an angle $\theta =\pi /4$ and the same values of $\beta$, i.e. $\beta =0.2$ (red lines), $\beta = 1$ (blue lines), $\beta =2$ (green lines). It can be observed a small jump in the phase velocity Fig 4.(a) for the case of $\beta$ $=$ 1, which is reflected also in the amplitude Fig 4.(d)

\begin{figure}[ht]
\includegraphics[width = 6.5 in, height= 6.5 in]{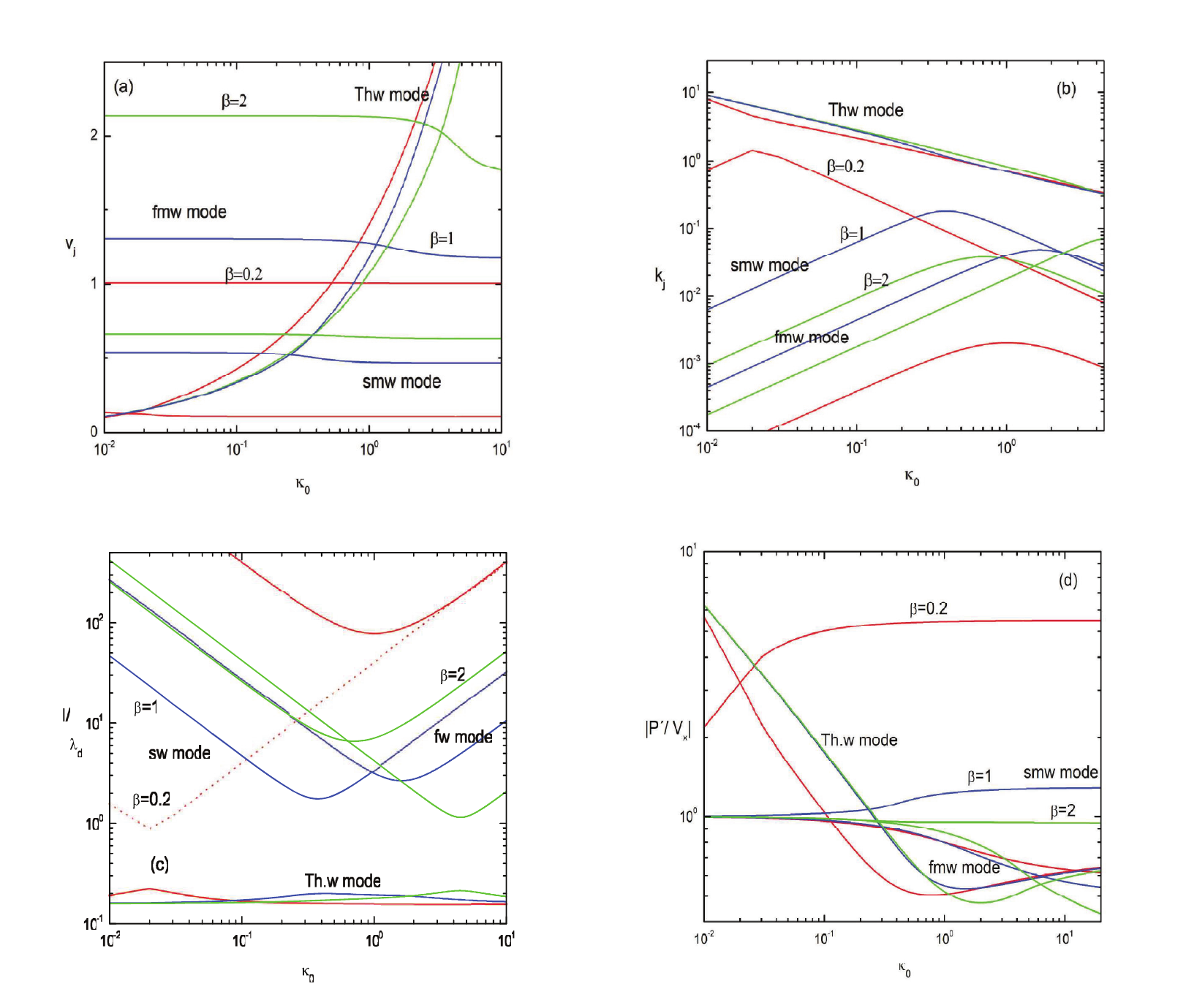}
\caption{Solution for the dispersion equation (\ref{madis}) using the thermal conductivity with $\mathbf{\theta =\pi/4}$. Here the phase velocity Fig 4.(a), the damping coefficient Fig 4.(b), the damping per unit wavelength Fig 4.(c) and the amplitude  $\left\vert \tilde{p}/\tilde{v}_{x}\right\vert$ Fig 4.(d) are plotted for the two slow and fast $mw$ modes, and the thermal $Thw$ mode with $\beta =0.2$ (red lines), $\beta=1$ (blue lines), and $\beta =2$ (green lines), it can be observed a small jump in the phase velocity 4.(a) for the case of $\beta$ $=$ 1, which is reflected also in the amplitude 4.(d).}
\end{figure}

For this particular value of $\mathbf{\theta =\pi/4}$, $\tilde{h}_{y}=\tilde{v}_{y}$ $=$ 0, and $\left\vert\tilde{p}/\tilde{v}_{x}\right\vert =\left\vert \left(1+\gamma^{-1}\tilde{\kappa} \tilde{k}^{3}\right)/\left(1+\tilde{\kappa}\tilde{k}^{2}\right)\right\vert$.

The thermal waves show a minimum of $\left\vert \tilde{p} /\tilde{v}_{x}\right\vert$ at the value of $\tilde{\kappa}_{0}$ at which the maximum damping per unit wave length $l_{d}/\lambda $ occurs. 

The magnetosonic wave showing its minimum of $l_{d}/\lambda$ at lower value of $\tilde{\kappa}_{0}$ presents an increasing value of $\left\vert \tilde{p} /\tilde{v}_{x}\right\vert $ when $\tilde{\kappa}_{0}$ increases, but the
magnetosonic wave with its minimum occurring at a larger value of $\tilde{\kappa}_{0}$ \ has a decreasing ratio $\left\vert \tilde{p}
/\tilde{v}_{x}\right\vert $ when $\tilde{\kappa}_{0}$ increases.

If only the magnetic diffusion ($\tilde{\nu}_{m}$) is accounted for, there should be no thermal waves because only the magnetic
terms are considered. In this case the dispersion equation (\ref{madis}) reduces to a quadratic equation for $\tilde{k}^{2}$ for which $|\tilde{p}/\tilde{v}_{x}|=1$. 

Furthermore, for $\theta =0$, a root becomes $\tilde{k} =1/\beta ^{2}$ , i.e. an undamped mode for which $h_{y}=0$ and $v_{y}=0$, and the another root becomes $\tilde{k}=\sqrt{i/(i+\tilde{\nu}_{m})}$. 

In Figs. 5 and 6 are shown the results for $\theta =\pi/4$ and 
$\frac{\pi}{2}$ respectively, and three different values of 
$\beta =$ 0.2 (red lines), $\beta =1$ (blue lines), $\beta =2$ (green lines). The the amplitude in these cases is $\left\vert\tilde{p}/\tilde{v}_{x}\right\vert=1$.

\begin{figure}[ht]
\includegraphics[width = 5.5 in, height= 5.0 in]{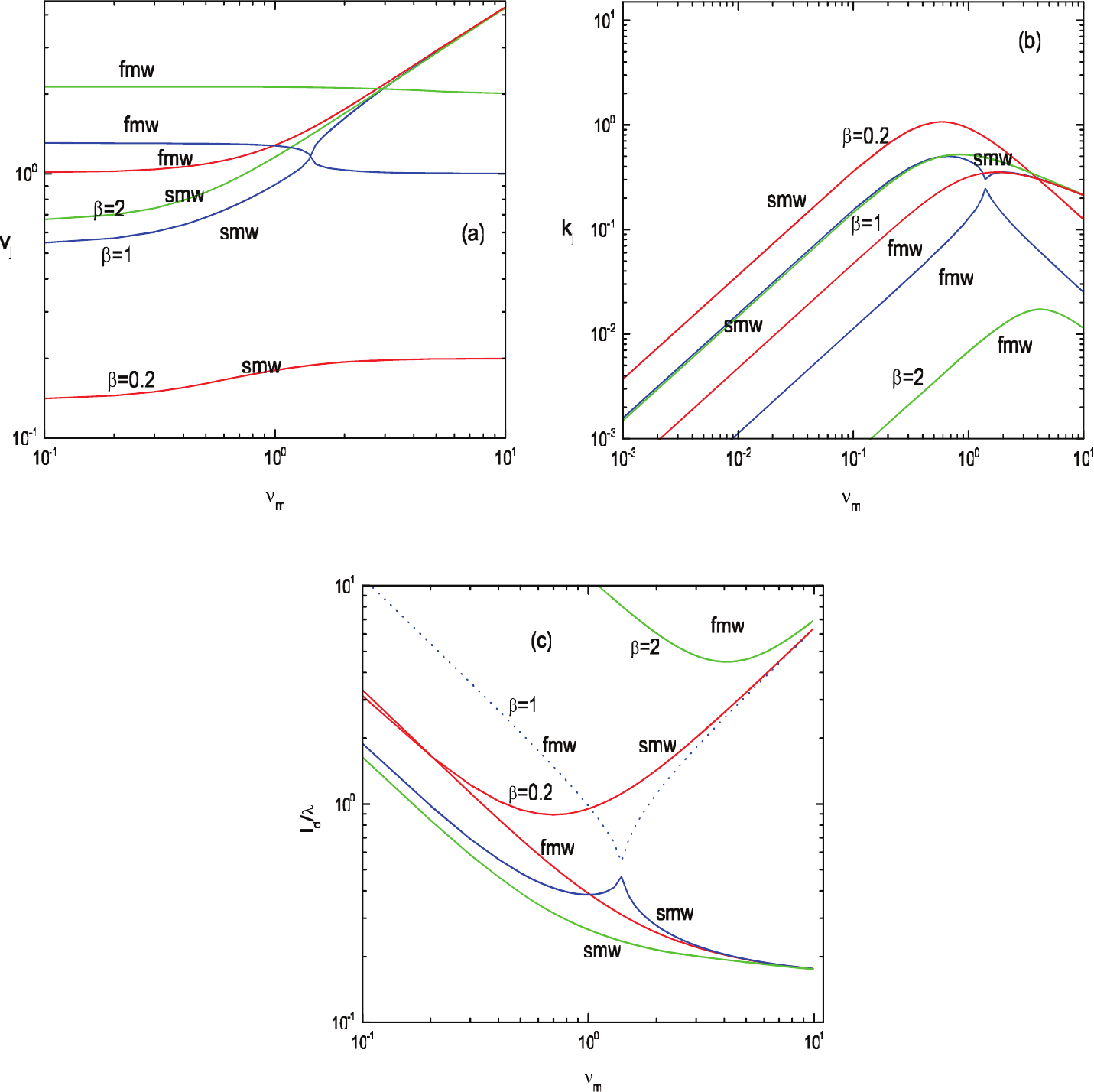}
\caption{The phase velocity - Fig 5.(a), the damping coefficient - Fig 5.(b), and the damping per unit wavelength - Fig 5.(c) for the magnetosonic fast and slow modes are plotted for $\beta =$ 0.2 (red lines), $\beta =1$ (blue lines), $\beta =2$ (green lines), for the dispersion equation (\ref{madis}) as function of the magnetic diffusivity with $\mathbf{\theta =\pi/4}$, for $\beta$ $=$ 1 (blue line), a crossing of slow and fast magnetosonic modes is observed.}
\end{figure}

When the magnetic energy density is of the order or larger than the kinetic energy in the wave $\beta$ $\leq$ 1, there is not
crossing of slow and fast modes, but mode crossing occurs when $\beta$ $=$ 1, see Figs 5a and 6a for instance. 

The damping coefficient for the slow mode is
a decreasing function of $\tilde{\nu}_{m}$ ($\sim \omega $) but that for the fast mode shows a maximum at a value of $\tilde{\nu}_{m}$ depending on the value of $\beta $ (Figs. 5b and 6b) and for the damping per wave length there is a corresponding minimum (Figs 5c and 6c). This minimum occurs at the mode crossing point when $\beta =1$. 

\begin{figure}[ht]
\includegraphics[width = 5.5 in, height= 5.0 in]{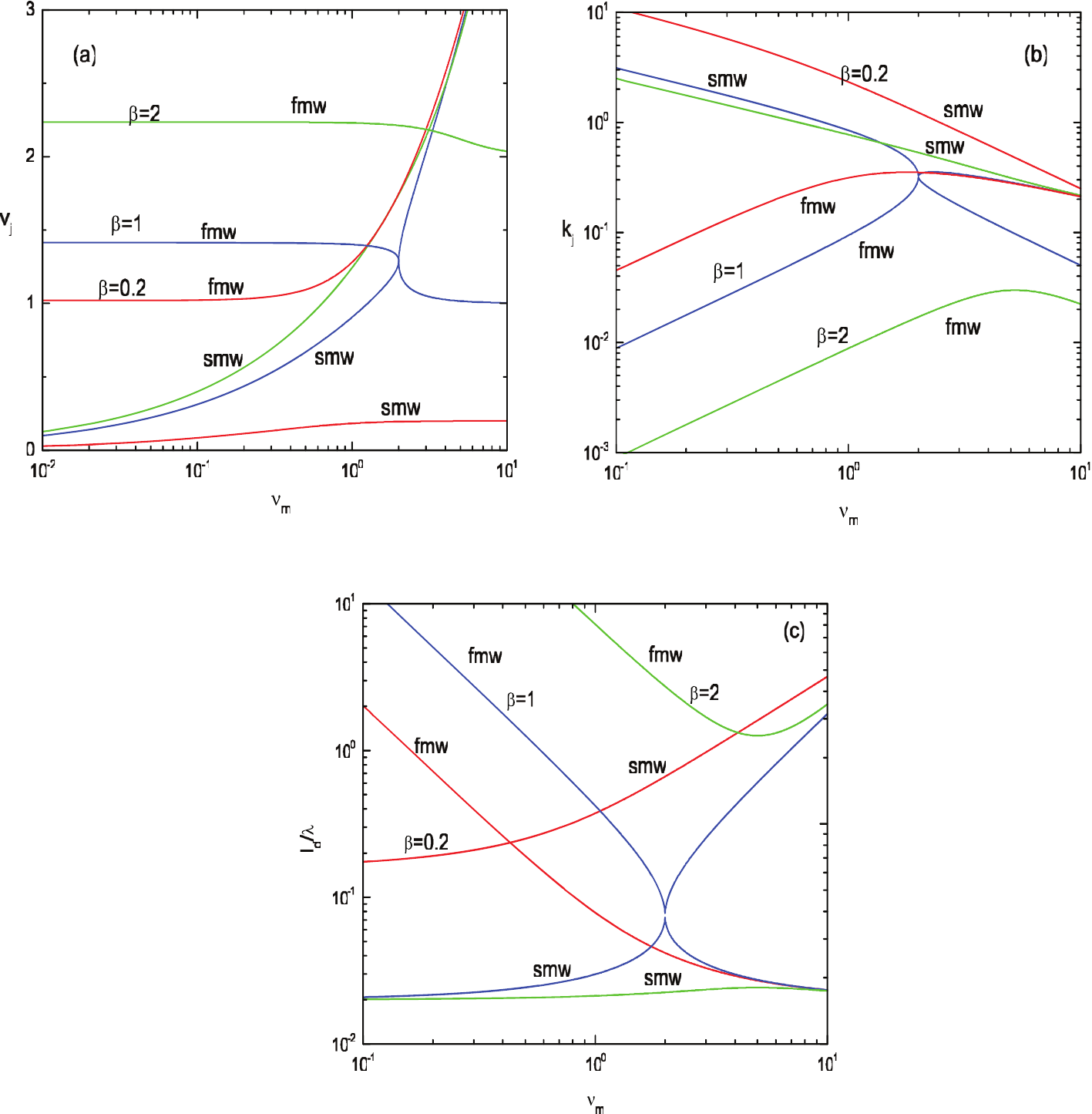}
\caption{The phase velocity - Fig 6.(a), the damping coefficient - Fig 6.(b), and the damping per unit wavelength - Fig 6.(c) for the magnetosonic fast and slow modes are plotted for $\beta =$ 0.2 (red lines), $\beta =1$ (blue lines), $\beta =2$ (green lines), for the dispersion equation (\ref{madis}) as function of the magnetic diffusivity with $\mathbf{\theta =\pi/2}$, for $\beta$ $=$ 1 (blue line), a crossing of slow and fast magnetosonic modes is observed.}
\end{figure}

Furthermore, for $\theta =\pi /2$, $ \tilde{v}_{y}=0$ and $|\tilde{h}_{y}/\tilde{v}_{x}|=\left\vert \tilde{k}/(1+i \; \tilde{\nu}_{m}\tilde{k}^{2})\right\vert$.

The case when only thermal conduction and heat/loss effects are accounted for in the equations, but neglecting the viscosities and the magnetic diffusion as well as the above asymptotic cases, but neglecting the anisotropy effects of the thermal conduction coefficient, have been analyzed in a previous work (\cite{Ib93}).

\clearpage

\subsection{Numerical analysis of the effect of the heat/loss function in the magnetosonic modes}

The case when in the energy equation the dissipative terms are neglected but the effects of the heat/loss are accounted for deserve further analysis, because this particular case is of great importance in many astrophysical as well as laboratory plasma. 

In this case the Eq.(\ref{madis}) reduces to
a quadratic equation in $\tilde{k}^{2}$ (if $\theta \neq \pi /2)$
corresponding to two magnetosonic waves modified by the heat loss input.

For $\theta =0$ one root becomes $\tilde{k}=1$ corresponding to an undamped Alfv\`{e}n wave for which $\left\vert \tilde{h}_{y}/\tilde{v}_{y}\right\vert=1$ and the other  root corresponding to the magnetosonic wave becomes 
$\tilde{k}=\sqrt{\gamma \left( i-\,\tilde{L}_{T}\right) \,/[i\,\gamma -\,\left( \tilde{L}_{T}-\tilde{L}\rho \right) ]}$.

For $\theta =\pi /2$, this is the only one root, but in this case, the magnetosonic wave has $\tilde{v}_{y}=0$.

As a first approximation, the heat/loss function can be can be parameterized to the form

\begin{equation}
L(\rho ,T)=\rho \phi_{i}(T)- C_{0}\, {\rho }^{a-1}\left( {\frac{T}{T_{i}}}
\right) ^{b},  \label{HCfunction}
\end{equation}

$\phi_{i}(T)$ being the piece-wise function $\phi_{i}(T)=\Lambda
\,_{i}\,\left(T/T_{i}\right)^{\eta}$, where $T_{i}$ and $\eta$ are parameters depending on the interval of temperature under consideration (see Table 1, \cite{Ve79}). 

Additionally, the parameters $C_{0}$, $a$ and $b$ depend on the heating processes considered. In particular:

\begin{enumerate}
    \item For a constant per unit volume heating $a=0$ and $b=0$.
    \item For a constant per unit mass heating heating $a=1$ and $b=0$.
    \item Heating by coronal current dissipation $a=1$ and $b=1$.
    \item Heating by Alfv\`{e}n mode/mode conversion $a=b=7/6$.
    \item Heating by Alfv\`{e}n mode/anomalous conduction damping $a=1/2$ and $ b=-a$. 
\end{enumerate}

See for instance  \cite{Ve79,Ro78} and references therein.

From Eq. (\ref{HCfunction}) follows that 
\begin{equation}
\tilde{L}_{\rho }(\rho ,T)={\frac{\left( 2-a\right) }{\eta -b}}\tilde{L}_{T},  \label{Lr}
\end{equation}

\begin{equation}
\tilde{L}_{T}(\rho ,T)=\left( \eta -b\right) \left( \frac{\rho \Lambda _{i}}{T_{i}c_{v}\omega }\right) 
\left( {\frac{T}{T_{i}}}\right) ^{\eta -1}. \label{Lt}
\end{equation}

The cooling function $\phi _{i}(T)$ has been plotted as a function of temperature in Fig. 7(a) in magenta color. 

\begin{figure}[ht]
\includegraphics[width = 6.0 in, height= 3.0 in]{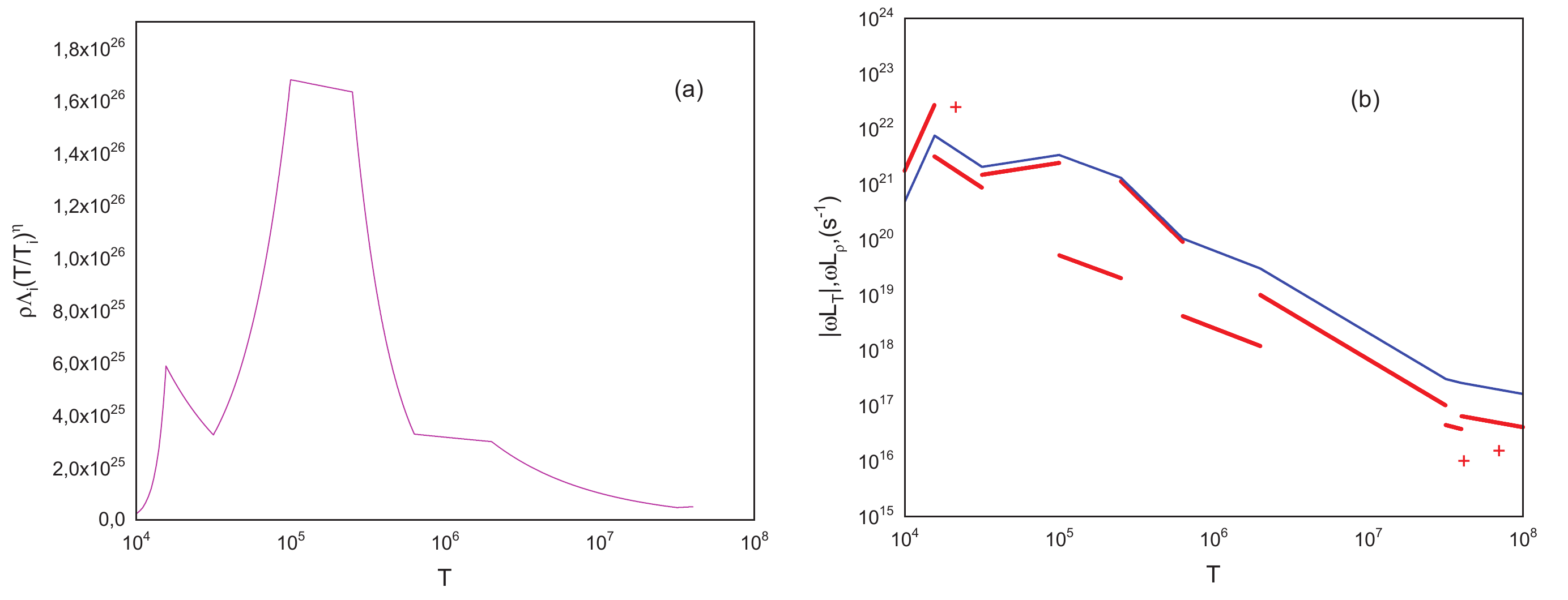}
\caption{For gases with solar abundances (a completely ionized gas ($\xi =1$) and a particle density $n= \frac{\rho N_{0}}{\mu}$ $=$ 1) the cooling function $\phi _{i}(T)$ has been plotted as a function of temperature in Fig. 7(a) in magenta color, the derivatives $\left\vert\omega \tilde{L}_{T}(\rho,T)\right\vert$ (red thick line) and $\omega \tilde{L}_{\rho }(\rho,T)$ (blue thin line) as function of $T$ have been plotted in Fig. 7(b) for a constant heating per unit volume (case 1)}
\end{figure}

The derivatives $\left\vert \omega \tilde{L}_{T}(\rho,T)\right\vert$ (red thick line) and $\omega \tilde{L}_{\rho }(\rho,T)$ (blue thin line) as function of $T$ have been plotted in Fig. 7(b) for a constant heating per unit volume (case 1), for a completely ionized gas ($\xi =1$) and a particle density $n= \frac{\rho N_{0}}{\mu}$ $=$ 1. 

The intervals of temperature where $\tilde{L}_{T}(\rho,T)$ $>$ 0 are indicated with the red label $+$, elsewhere 
$\tilde{L}_{T}(\rho ,T)$ $<$ 0.

The plots corresponding to the cases (2) to (5) also are shown: Fig. 8(a) for a constant per unit mass heating heating, fig. 8(b)
shows the heating by coronal current dissipation, fig. 8(c) plots the heating by Alfv\`{e}n mode/mode conversion, the heating by Alfv\`{e}n mode/anomalous conduction damping is shown in fig. 8(d). 

\begin{figure}[htbp!]
\includegraphics[width = 5.5 in, height= 5.0 in]{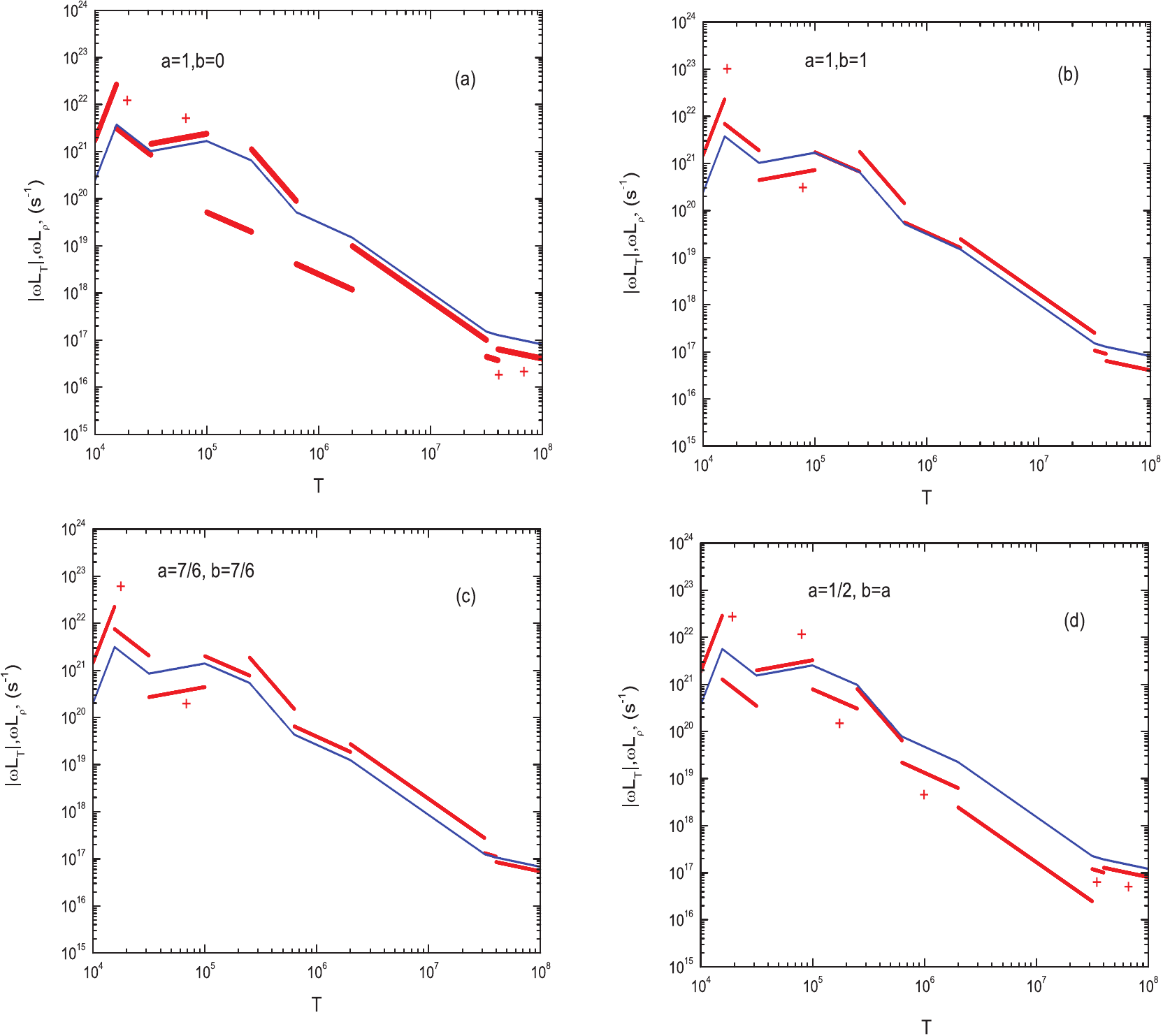}
\caption{The heat/loss function derivatives $\left\vert \omega \tilde{L}_{T}(\rho,T)\right\vert$ (red thick line) and $\omega \tilde{L}_{\rho }(\rho,T)$ (blue thin line) as function of $T$, that correspond to cases (2) to (5) are shown in figures (a) to (d) for a completely ionized gas ($\xi = 1$) and a particle density $n \frac{\rho N_{0}}{\mu}$ $=$ 1}
\end{figure}

Due to the fact that the cooling term in Eq.(\ref{HCfunction}) as well as its derivatives respect to temperature and density are $\sim$ $\rho$, for other densities, the corresponding values simply must be multiply by the factor $n$.

\clearpage

\section{Kinetic coefficients for a Hydrogen ionization plasma}

At this Section the kinetic/dissipation coefficients in a magnetic field, for the case of a recombining hydrogen plasma will be quoted out and briefly discussed. 

According (\cite{Sp62,Br65,La87,La60,Pi81}), for a hydrogen gas with ionization $\xi$ the two electric  conductivity  tensors are respectively given by

\begin{equation}
\sigma _{\perp }= 6.97\times 10^{7}\frac{T^{3/2}}{\ln \Lambda}, \label{eleccond} 
\end{equation}
and

\[
\sigma_{\parallel}=1.96\sigma_{\perp };
\]

The thermal conduction coefficients are expressed as

\begin{equation}
\kappa _{\parallel }=2.50\times 10^{3}(1-\xi )T^{1/2}+1.84\times 10^{-5} \frac{\xi T^{5/2}}{\ln \Lambda},  \label{Kapapa}
\end{equation}
and 

\begin{equation}
\kappa _{\perp }=1.48\times 10^{-17}\frac{\xi ^{2}n^{2}}{H^{2}T^{1/2}}\quad, \label{Kapape}
\end{equation}

Finally, the kinematic viscosity coefficient is given by

\begin{equation}
\eta =2.21\times 10^{-15}\frac{T^{5/2}}{\ln \Lambda }, \label{eta}
\end{equation}
and the kinematic viscosity is expressed as (\cite{La87})

\begin{equation}
\nu = \frac{\eta}{\rho}. \label{zeta}
\end{equation}

The logarithmic coefficients $\ln \Lambda$ are for temperatures $T < 4.2 \times 10^{5}K$
\[
\ln \Lambda =
23.24 + \ln \left(\frac{(10^{-4}T)^{3}}{n\xi}\right) ^{1/2}
\]
or when the temperature $T > 4.2 \times 10^{5}K$
\[
\ln \Lambda = 
29.71 + \ln \left( \frac{10^{-6}T}{(n\xi )^{1/2}}\right).
\]

On the other hand and as a first approximation the total dissipative
coefficient for magnetosonic waves can be writing as 

\begin{equation}
\gamma _{d} \approx \frac{4}{3}\nu +(\gamma -1)\chi +\nu_{m},  \label{totaldisco}
\end{equation}

where $\chi =\kappa/\rho c_{p}$ is the thermometric conductivity and $\nu_{m}$ the magnetic diffusion (\cite{Pi81,La87,Co10}).

Note that at the present approximation $\nu(T)$, $\nu_{m}(T)$, and $\chi(n,T,\xi)$ explicitly depend on the particular form of the rate function $X(n,T,\xi)$ and the wave frequency (\cite{Ib04,Co10,Ib19}). In (\cite{Ib19}) the problem of reacting gases and the bulk viscosity has been discussed to some extent.

In Fig 9(a) the quantities $4\nu/3$ (black solid line) and $(\gamma -1)\chi$ have been plotted as functions on temperature for $n$ $=1$ and four values of the ionization $\xi =10^{-6}$ (blue colour), $10^{-3}$ (red colour), $10^{-1}$(brown colour) , and \ $0,99$(green colour). 

Note that  $\chi\sim n^{2}$, therefore, the effect of increasing (decreasing) the density is to increase (decrease) the respective values of $\chi$. The value of $\nu_{m}$ $\ll$ $10^{10}$ $cm^{-2}$ $s^{-1}$ in the range of T under consideration has not been plotted.

However, $\nu_{m}$ parallel and perpendicular to the magnetic field can become of the order or greater than of $\frac{4}{3}\nu$ and $(\gamma -1) \chi$ for high densities ($n \geq 10^{10}$ $cm^{-3}$) 
and strong magnetic fields $H\geq 1$ G, say for instance in the solar low atmosphere and photosphere. 

For context, in Fig. 9(b) all dissipation coefficients are shown for $H=1$ G and $n=10^{15}$cm$^{-3}$; from where it is apparent that the magnetic dissipation parallel ($\nu_{m}$) as well as perpendicular ($\nu _{m\perp}$) to the magnetic field becomes dominant in range of temperatures depending on the particular values of the ionization degree as well as the particle density.

In Fig 9(b) the quantities $4\nu/3$ (black solid line) and $(\gamma -1)\chi$ have been plotted as functions on temperature for $n$ $=1$ and four values of the ionization $\xi =10^{-6}$ (blue colour), $10^{-3}$ (red colour), $10^{-1}$(brown colour) , and $0,99$ (green colour). 

The perpendicular magnetic diffusion ($\nu_{m\perp}$)  is plotted is Fig 9(b) for four values of the ionization $\xi =10^{-6}$ (gray point line), $10^{-3}$ (black point line), $10^{-1}$(brown point line), and $0,99$ (magenta point line).

The parallel magnetic diffusion ($\nu _{m\parallel}$) is also plotted is Fig 9(b) (dash black line). 

\begin{figure}[ht]
\includegraphics[width = 6.5 in, height= 3.5 in]{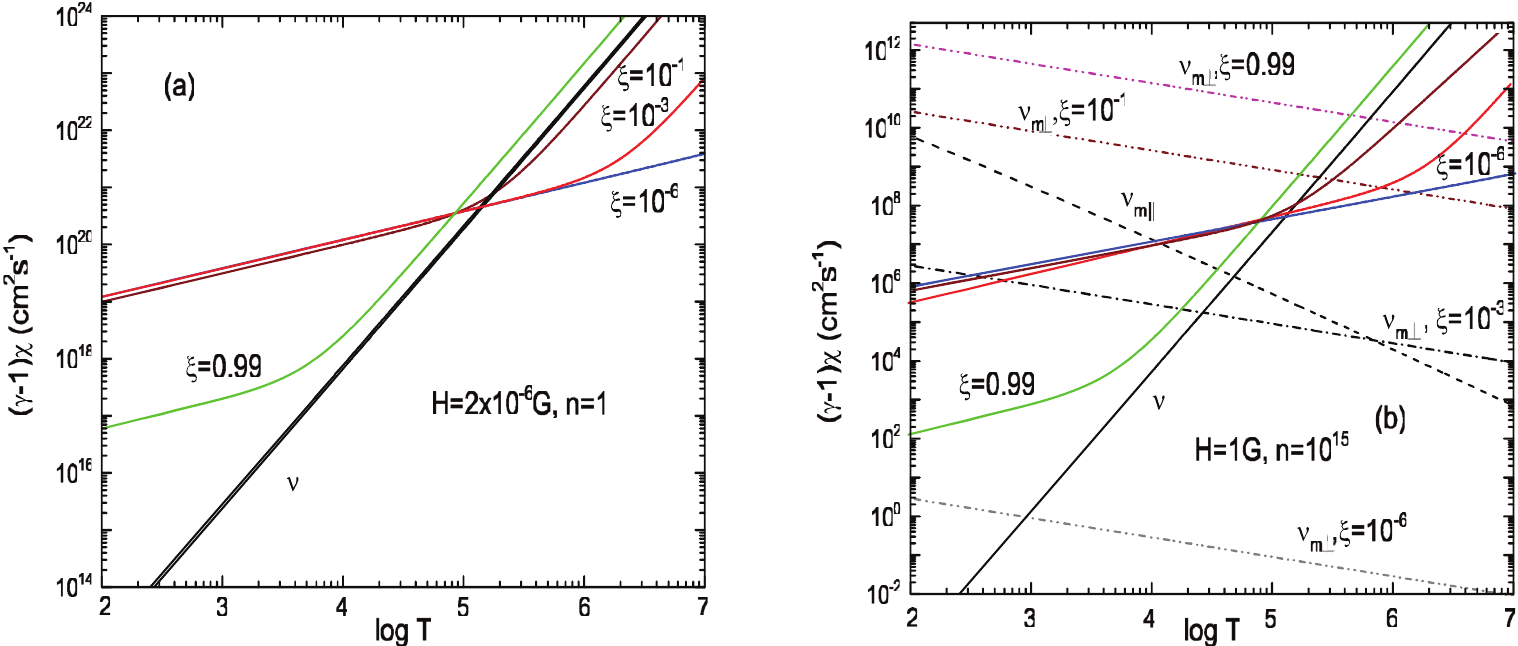}
\caption{In Fig 9. (a) (without magnetic diffusion)  and (b) (with ($\nu_{m}$)), the quantities $4\nu/3$ (black solid line) and $(\gamma -1)\chi$ have been plotted as functions on temperature for four values of the ionization $\xi =10^{-6}$ (blue colour), $10^{-3}$ (red colour), $10^{-1}$(brown colour) , and $0,99$ (green colour).}
\end{figure}

\clearpage

\section{Conclusions}
The present work was aimed at investigating the behavior and propagation of MHD waves in optically thin plasmas, with ionization and dissipative effects. The results are summarized in four sections. 

In section II, the set of MHD equations was linearized, leading to two independent cases where each matrix generates a dispersion relation whose roots for the case of Alfv\`{e}n waves are a complex equation, this relationship led to the deduction of the Landau damping expression in a different way to the one presented in classical texts.

In section III, for the linear approximation it was observed that both, thermal and magneto-acoustic modes  are damped by the thermal conduction, viscosity and the influence of the cooling-heating function. The complex eigen-equation was described with some detail, and several asymptotic cases of the full polynomial solutions were discussed (\ref{madis}):
\begin{itemize}
    \item  The case when the only dissipative process taken into account is the thermal conductivity was discussed for several values of $\theta$ in Eq.(\ref{madis}). We found eigenvalues corresponding to two damped magnetosonic waves, and a thermal wave. We also found a small jump in the phase velocity for magnetosonic modes for the case of $\beta = 1$ which is reflected also in the amplitude.
    \item In the case with only the magnetic diffusion term ($\tilde{\nu}_{m}$), the dispersion equation (\ref{madis}) reduces to a quadratic equation for $\tilde{k}^{2}$ for which $|\tilde{p}/\tilde{v}_{x}|=1$ and lacks the thermal mode. It was found that if the magnetic energy density is of the order or larger than the kinetic energy in the wave for $\beta$ $=$ 1, a crossing of slow and fast magnetosonic modes was observed.
\end{itemize}

In section IV, in the energy equation the dissipative terms were neglected, but the effects of the heat/loss were accounted, because of its great importance in many astrophysical as well as laboratory plasma applications. In this case the Eq.(\ref{madis}) reduces to a quadratic equation in $\tilde{k}^{2}$ corresponding to two magnetosonic waves modified by the heat loss input. We
described in this section five heating processes for a thin optical plasma.

Finally, in section V, the kinetic coefficients in a magnetic field, for the case of a recombining hydrogen plasma were briefly discussed. It was found that the magnetic dissipation parallel ($\nu_{m}$) as well as perpendicular ($\nu_{m\perp }$) to the magnetic field become dominant in a range of temperatures depending on the particular values of the ionization degree $\xi$ as well as the particle density $n$.

\section*{Acknowledgments}

This work is dedicated to the memory of Prof. Miguel H. Ibañez S.
(1945 - 2020) who devoted his life to the theoretical study of astrophysical plasmas and fluid dynamics in reacting gases, and non-adiabatic flows. 

He always will be remembered as a devoted researcher, a teacher of several generations of Venezuelan physicists, and a unique friend.

This research did not receive any grant from either the University of Los Andes or a government organization.

\bibliography{mybibliography}

\end{document}